\documentstyle[12pt,aaspp,psfig]{article}

\def\ltsima{$\; \buildrel < \over \sim \;$}
\def\lsim{\lower.5ex\hbox{\ltsima}}

\begin{document}
\title
{Evidence for a source size of less than 2000 AU in Quasar 2237+0305}

\author{Atsunori Yonehara\altaffilmark{1,2,3}}

\altaffiltext{1}{Department of Astronomy, Kyoto University, Sakyo-ku, 
 Kyoto 606-8502, Japan}
\altaffiltext{2}{e-mail: yonehara@kusastro.kyoto-u.ac.jp}
\altaffiltext{3}{Research Fellow of the Japan Society 
 for the Promotion of Science}

\begin{abstract}
Recently, OGLE team have reported clear 
 quasar microlensing signal in Q2237+0305.
We have analyzed the microlens event of ``image C''
 by using their finely and densely sampled lightcurves.
From lightcurve fitting, we can unambiguously set 
 the source size of $\lsim 0.98$ Einstein Ring radius
 as a conservative limit.
This limit corresponds to $2000 {\rm (AU)}$, if we adopt 
 $M_{\rm lens} \sim 0.1M_{\odot}$ obtained by 
 a recent statistical study of mean mass of lens object.
This gives a clear evidence for the existence of an accretion disk 
 in the central region of the quasar.

\end{abstract}


\section{INTRODUCTION}

It is widely believed that central origins driving activity 
 of quasars and AGNs (Active Galactic Nuclei) is 
 an accretion disk surrounding a $10^{6 \sim 9} M_{\odot}$ 
 SMBH (supermassive black hole).
To find a direct evidence for this general belief is 
 one of the most exciting subjects in the current research 
 in astronomy and astrophysics, but unfortunately, 
 the expected angular size of accretion disks is too small ($\lsim 1~\mu as$)
 to directly resolve spatially by using present observational instruments.
For this reason, a proof of SMBH hypothesis remains as an unsolved problem
 in fields of quasar/AGN. 
Such a situation will not alter in near future.

However, there is a strong tool to make it possible.
That is so-called ``quasar microlensing''.
Following the first report of a detection of quasar microlensing 
 in Q2237+0305 (``Einstein Cross'' or ``Huchra's lens'')
 by Irwin et al. (1989) and subsequent extensive theoretical works 
 (e.g., Wambsganss 1990), 
 many researchers focused on this interesting 
 subject and presented many meaningful results. 

Roughly speaking, there are two different approaches 
 to probe a structure of the central region of quasars.
One is statistical approach by using long term monitoring data
 (e.g., {\O}stensen et al. 1996) 
 that are expected to contain many microlens events.
Recently, Wyithe et al. have performed thorough statistical study;
 they compare real and mock observational results, 
 and constrain transverse velocities of lens, the mean mass of the lens, 
 source size, and so on (Wyithe, Webster, \& Turner 1999, 2000b, 2000c).
Another is to focus on a single HME (High Magnification Event), 
 on the basis of reasonable assumptions and/or some statistical features, 
 and constrain a source size or structure 
 (e.g., Wyithe et al. 2000a, and references therein).

In this {\it letter}, thus we take the latter approach and  
 performed lightcurve fitting for the accurately and densely observed 
 microlensing event that OGLE team have detected in Q2237+0305 lately
 (Wozniak et al. 2000a, 2000b), 
to put a limit on the microlensed source size.
In section 2, we briefly present our method to fit an observed lightcurve
 and the results and discussions are shown in section 3.

\section{LIGHTCURVE FITTING}

The {\it V}-band monitoring data of OGLE team show that
 there is a dramatic brightening in image A and 
 abrupt brightening and subsequent decay in image C ($> 0.5 {\rm mag}$), 
 whereas flux changes have been less significant in images B and D 
 (they have only $\sim 20$ \% flux variation) compared with image A and C.
This suggest that microlens events have occurred 
 in images A and C independently.
While the behavior of image A was complicated
 and could be caused by complicated (many) caustics, 
 that of image C seems to be quite simple and 
 can be understood in terms of a single-caustic induced HME.
Therefore, applying approximate magnification formulae  
 which are appropriate in the vicinity of a caustic, 
 we try to fit the lightcurve of image C.
Concrete formulae of magnification ($\mu$) for a source position $(x, y)$
 are shown in Fluke \& Webster (1999) for
 ``fold caustic (hereafter, FC)'' case
 and Zakharov (1995) for ``cusp caustic (hereafter, CC)'' case.

To characterize properties of caustics, e.g., curvature of fold caustic, 
 we should evaluate derivatives of Fermat potential ($\phi$)
 for gravitational lensing phenomena.
In quasar microlensing cases, Fermat potential is written as follows, 
\begin{eqnarray}
\phi &=& \frac{1}{2} \{ (\eta -x)^2 + (\xi -y)^2 - \kappa_{\rm c} 
 (\eta^2 +\xi^2) - \gamma \left[ ( \eta^2 -\xi^2 ) \cos 2 \theta 
  + 2 \eta \xi \sin 2\theta \right] \} \nonumber \\
 & ~ &  - \sum_{i=1}^{N_{\rm lens}} 
  \log \{ \epsilon_i \left[ (\eta -\eta_i)^2 +(\xi -\xi_i)^2 \right]^{1/2} \}, 
\label{eq:lenspot}
\end{eqnarray}
where, $(\eta, \xi)$ is an image position, 
$\gamma$ and $\theta$ is shear and its direction, 
 $\kappa_{\rm c}$ is convergence arising due to
 continuous (smooth) mass distribution,
 $(\eta_i, \xi_i)$ and $\epsilon_i$ in this equation
 represent the $i$-th lens position and its normalized mass,
 and $N_{\rm lens}$ is the total number of compact, stellar lens.
Here, $\kappa_{\rm c}$ plus contributions of
 compact, stellar lens should be almost equal to 
 total convergence ($\kappa$) that is determined by macrolens model.
All the length scales are normalized by the Einstein-ring radius ($r_{\rm E}$).

Unfortunately, we are not able to know complete spatial distribution 
 and mass of compact, stellar lens objects, and 
  it is impossible to characterize exact properties of caustics.
Therefore, in this study, we assume that only one single lens object plays 
 a significant role in the HME and 
 contributions from other lens objects are negligible ($N_{\rm lens} = 1$), 
 i.e., we only consider about so-called 
 ``Chang-Refsdal lens (Chang \& Refsdal 1984)'' situation 
 (single lens object plus external convergence and shear), 
 for simplicity, and characterize the properties.
Applied values of convergence and shear is 
 one of the best fit value on image C for macrolens model of Q2237+0305, 
 $\kappa = 0.69$ and $\gamma = 0.71$ (Schmidt, Webster \& Lewis, 1998).
In some complicated case (e.g., lens objects are clustering), 
 situation may be modified, or in the worst case, 
 above assumption will break down. 
However, if this assumption holds (e.g., $\kappa _{\rm c} \sim \kappa$
 is the case), we are able to 
 characterize the feature of FC and CC by only one parameter ($\theta$).

In the case of the Chang-Refsdal lens, 2 different types of cusp caustic 
 will be formed at 4 or 6 angular positions of the closed caustic curve 
 (see Chang \& Refsdal 1984), and the fold caustics 
 appear as connecting these cusp caustics. 
Thus, we consider 2 cusp cases and 4 representative fold cases 
 for an angle range of $\theta = 0 \sim \pi/2$ because of its symmetry. 
Moreover, we only consider about circular-shape source 
 with top-hat brightness profile,
 i.e., neglect the effect of inclination angle, brightness profile.
Therefore, magnification (${\cal M}$) at any given source position $(x, y)$ 
 is obtained by the integration of $\mu$ on the circular disk with radius $R$.
Assuming top-hat brightness profile is fairly simple treatment 
 because the shape of microlensing lightcurve depends on 
 source brightness profile (e.g., Yonehara et al. 1999).
However, the resultant source size can be regarded as
 an effective (equivalent) source size and we use 
 the top-hat brightness profile for convenience to numerical integrations
 (with $1000$ mesh number in this study).

To include magnification caused by macrolensing and 
 another caustics for microlensing, 
 here, we add constant magnification, $M_{\rm 0}$, plus 
 gradual change of the magnification, $\dot{M_{\rm 0}} \cdot t$, 
 to the total magnification.  
The latter mimics magnification changes for the ensemble of
 other microlens and this gradual change make the fits better 
 (Wyithe and Turner, private communication).
Furthermore, intrinsic variability of this quasar may be 
 long duration with small amplitude (Wyithe et al. 2000d),
 and this term may also mimic the intrinsic variabilities of quasar.

Additionally, we should evaluate the apparent magnitude of quasar
 without any microlensing and macrolensing,
 i.e., the intrinsic magnitude, $m_{\rm 0}$.
It is quite difficult because quasars have intrinsic variabilities and
 also the observed magnitude of Q2237+0305 is affected by
 small amplitude and/or long timescale microlensing.
Fortunately, from the monitoring data by {\O}stensen et al.(1996) and 
 the OGLE team, the observed magnitude of image C is roughly constant 
 at $\sim 18.6 ~{\rm mag}$ long before the recent HME.
We thus take this as the magnitude of this quasar without any microlensing.
Moreover, by using the previously applied value of $\kappa$ and $\gamma$,
 we can evaluate the macrolens magnification of image C, 
 and the intrinsic magnitude is estimated to be
 $m_{\rm 0} = 18.6-2.5\log \left[ (1-\kappa)^2 - \gamma^2 \right] \sim 19.6$.

Finally, assuming the source trajectory is straight and
 determine source velocity on the source plane,
 $\vec{v} = (v_{\rm x}, v_{\rm y})$, time when the source crosses a caustic
 (FC case) or $x$-axis (CC case), $T_{\rm 0}$, 
 the expected microlensing lightcurve for any given parameter is 
 obtained from  
 $m(t) = m_{\rm 0} - 2.5\log \left[ M_{\rm 0} + \dot{M}_{\rm 0} \cdot t 
 + {\cal M}(t) \right]$
 (for CC case and fold caustic gazing case, 
 impact parameter, $d$ is also required).

To obtain the best fit lightcurve and its parameter, 
 we minimize $\chi^2$ value between the observed lightcurve, $m_{\rm obs}(t)$,
 and the mock lightcurve for a given parameter, $m(t)$, 
 for each caustic case (FC and CC) by using one of downhill simplex method, 
 so-called ``AMOEBA'' routine (Press et al. 1986). 
Nonetheless, caustic is a kind of singularities, 
 and fitting methods including singularities do not work so well.
For this reason, we subdivide each caustic case into 
 all considerable path cases (depicted in figure~\ref{fig:sche})  
 and perform lightcurve fitting at every possible case.
After the best fit parameters are obtained, we compare reduce $\chi^2$ for
 all the possible cases for each of FC and CC, and determine 
 the best fit (smallest $\chi^2$) parameters for each FC and CC case.

In this lightcurve fitting, we only took into account data points
 around the peak of image C ($JD-2450000 = 1289.905 \sim 1529.531$,
 total number of used data points is $83$).
Of course, we can also taken into account data points 
 before and after the peak, but in those epochs, 
 distance between the source and caustic
 could be larger than that in the peak region.
Thus, it is not clear whether the approximation for magnification 
 is reasonable or not before and after the peak.
And so, we have restricted data points only around the peak.  

\section{RESULTS AND DISCUSSIONS}

\begin{figure}[htbp]
\centerline{\psfig{figure=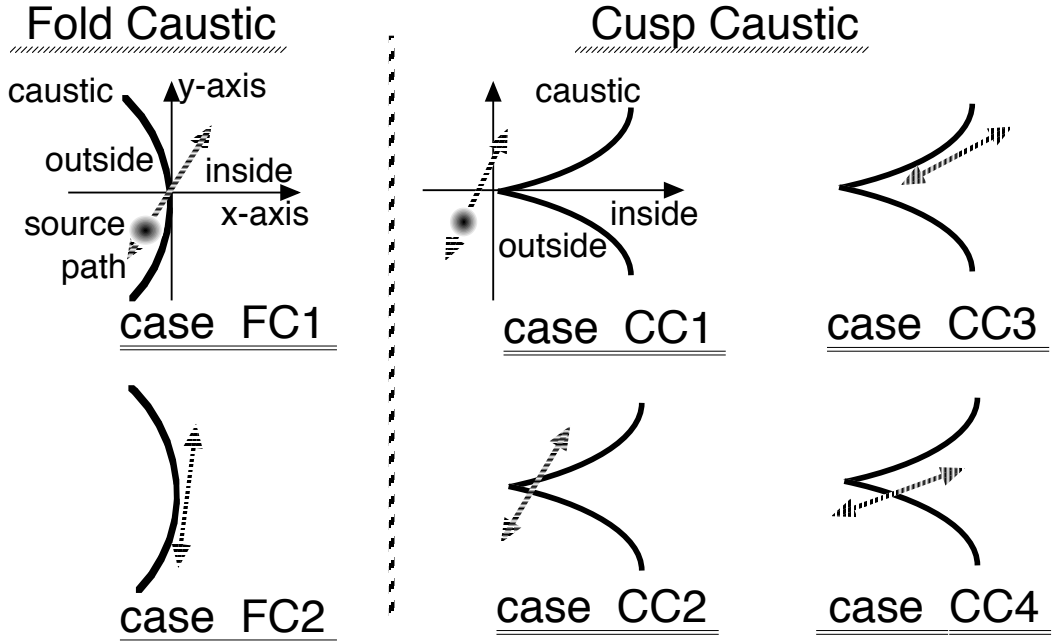,height=9cm}}
\caption{Schematic view of possible HME. 
 All the six cases to reproduce the observed, fairly symmetric HME event 
 are shown: 2 are fold caustic cases and 4 are cusp caustic.
}
\label{fig:sche}
\end{figure}

\begin{figure}[htbp]
\centerline{\psfig{figure=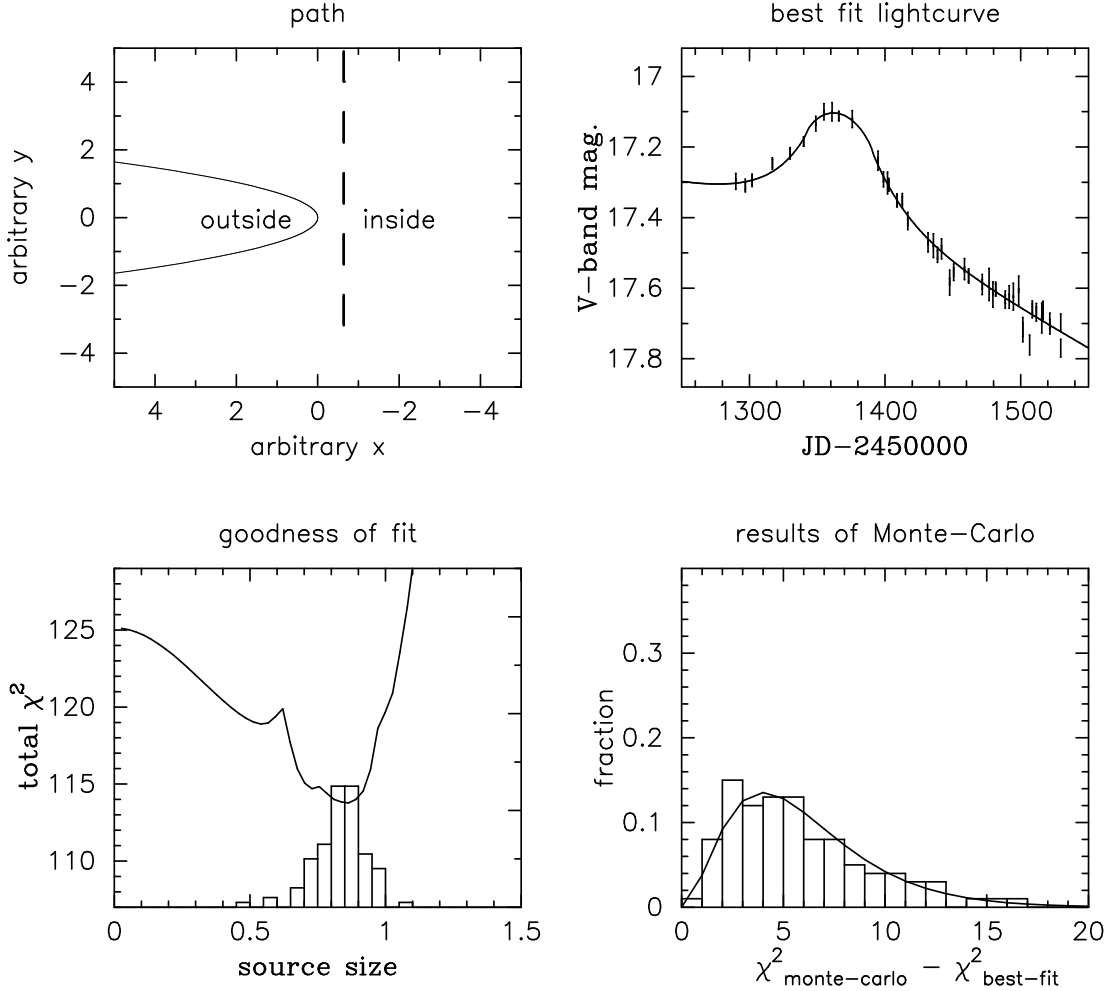,height=13cm}}
\caption{Fitting results in a case of fold caustic (fold-3).
In upper left panel, the best fit source path (dashed line) and
 relative to caustic (bold line) is depicted.
Upper right panel is observed flux (error bar) and 
 the best fit lightcurve (bold line).
Lower left panel is total $\chi^2$ value respect to source size (bold line), 
 and result of Monte-Carlo simulation (histogram).
Lower right panel  distribution of $\chi^2$ with 
 5 degree of freedom (bold line)
 and distribution of $\chi^2$ difference between mock lightcurve 
 and the best fit lightcurve (histogram). 
All the length scale is $r_{\rm E}$.
Kinks in source size dependence of total $\chi ^2$ (lower left panel)
 are caused by changes from a best-fit subdivided case
 to another subdivided case, e.g., at source size $ \sim 0.6$, 
 the best-fit, subdivided case changes from FC1 to FC2. 
}
\label{fig:resfold}
\end{figure}

\begin{figure}[htbp]
\centerline{\psfig{figure=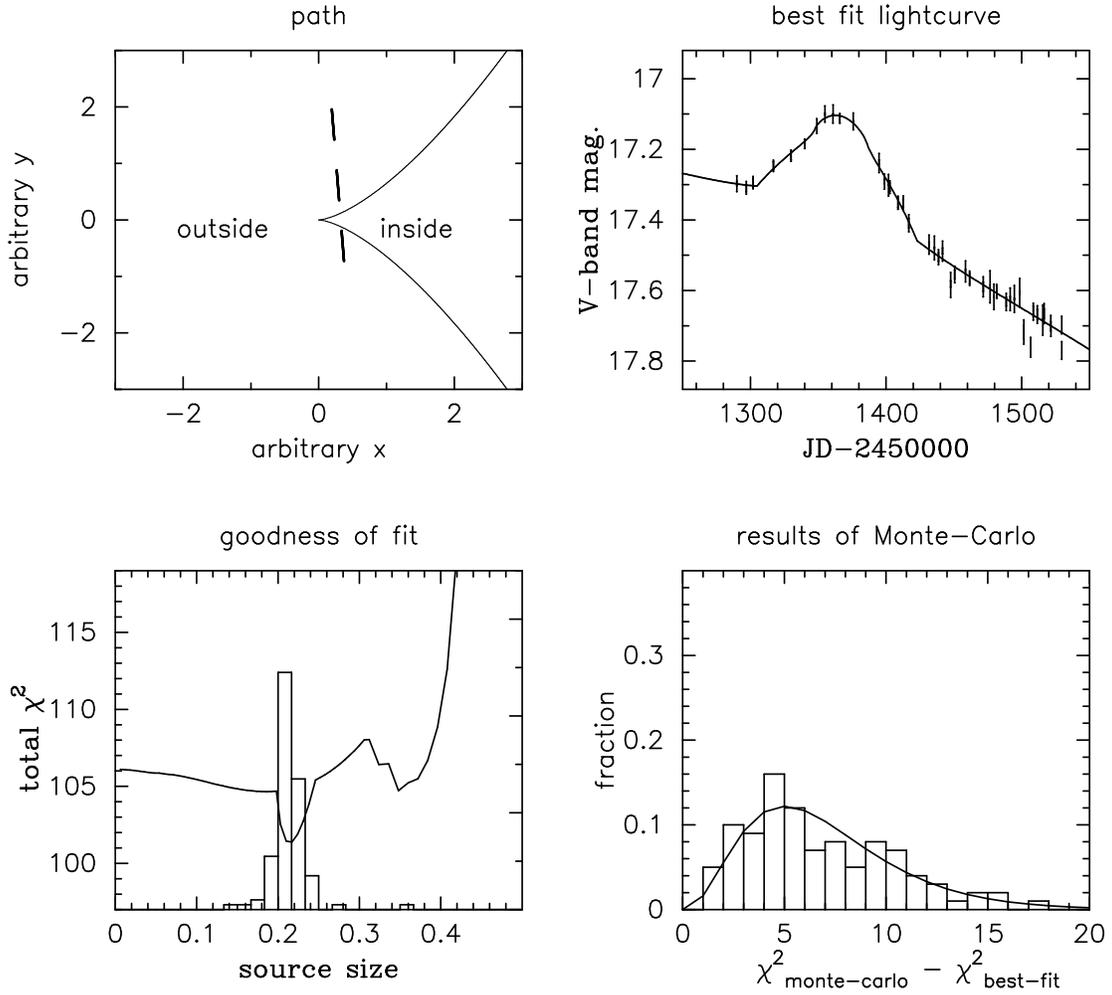,height=13cm}}
\caption{Same as figure~\ref{fig:resfold} but cusp caustic case (cusp-1).
Degree of freedom of lower right panel is 6 in this case.
Kinks in source size dependence of total $\chi ^2$ value
 (lower left panel) appeared 
 with the same reason as in figure~\ref{fig:resfold} 
}
\label{fig:rescusp}
\end{figure}

Our resultant, best fit parameters for all the cases 
 that we have considered are summarized in table~\ref{tab:wp}.
Some resultant path of source, the best fit lightcurve, 
 source size dependence of reduced $\chi^2$ are 
 also shown in figures~\ref{fig:resfold} and~\ref{fig:rescusp}
(degree of freedom is $83-5 = 78$ for FC, $83-6 = 77$ for CC).
In the case of a infinitely-small size source, 
 expected lightcurves for microlens event are 
 quite different from case to case, i.e., 
 a lightcurve for FC case and that of CC case is clearly different,
 and also, difference between that of fold-1 and fold-2
 (figure~\ref{fig:sche}) is evident and so on.
But, as you can easily see in these table and figures,
 the fitting for all cases works well 
 (the best-fit reduced $\chi^2$ is $\sim 1$).
This fact owing to the finite-size source effect, 
 and we can manage to reproduce the observed feature
 at every considerable case.

We also performed a Monte-Carlo simulation for every case to estimate
 confidence region of best fit parameter by using following procedures; 
(1)~Supposing that the best fit parameter is real parameter, 
 we calculate an ideal lightcurve without any errors, 
(2)~Add random errors with the magnitude corresponding to 
 the observational error dispersion, and sample this lightcurve 
 at the times corresponding to the actually observed times,
 we obtain a mock lightcurve. 
(3)~By using this mock lightcurve, 
 we perform a lightcurve fitting again for all considerable cases 
 (at FC and CC case indicated before, see also fig~\ref{fig:sche})
 and obtain a set of the best fit parameters for the mock lightcurve.  

Iterating procedure (2) and (3) for 100 times in this study, 
 summarizing the best fit values for mock lightcurves, 
 and we can evaluate a confidence region.
To evaluate the $90\%$ confidence region from Monte-Carlo results,  
 we calculate total $\chi^2$ between lightcurve which is actually observed 
 and that is obtained from the parameters of Monte-Carlo result. 
Subsequently, we pick up $90\%$ parameter sets which have
 smaller total $\chi^2$. 
From these selected parameter sets, finally, we can obtain
 maximum and minimum values of parameters and we define
 ranges between these maximum and minimum as the $90\%$ confidence region.

In figures~\ref{fig:resfold} and~\ref{fig:rescusp}, 
 we also presented a histogram for the $\chi^2$ differences
 between ``mock'' lightcurves and the best fit lightcurve,
 and ideal $\chi^2$ distribution curves for corresponding degrees of freedom 
 (5 for FC, 6 for CC).
These two exhibit similar distributions and 
 our confidence region estimate seems to be reasonable.

For every case, if the source size is larger than the best fit value, 
 expected magnification will be suppressed, lightcurve will become shallow,
 and goodness of fit is reduced.
On the other hands, if the source is smaller than the best fit value, 
 expected magnification will be enhanced, lightcurve will become sharp, 
 and goodness of fit reduced, too.
These are qualitative reason  
 why the source size is limited in somewhat small range.

Considering all our fitting result, at least, 
 we can say that the source size of Q2237+0305 should be smaller than 
 $\sim 0.98$ Einstein-ring radius (more than $90\%$ confidence level).
This upper limit is given in the case of FC (fold 3 in table~\ref{tab:wp}), 
 while another case suggests a much smaller source size.
Thus, this limit is a fairly conservative upper limit for the source.
Since the Einstein-ring radius of quasar microlensing is typically 
$\lsim 1 ~\mu{\rm as}$, 
our result indicates the existence of a sub-$\mu{\rm as}$ source in quasar !

To obtain the actual size, 
 we have to calculate $r_{\rm E}$ for relevant parameters. 
If we assume $1.0 M_{\odot}$ as a mass of lens object ($M_{\rm lens}$) and 
 the Hubble constant $H_{\rm 0} = 67 {\rm km~s^{-1}~Mpc^{-1}}$ 
 (Kundi\'c et al. 1997), $r_{\rm E}$ will be corresponds to $10^{17} {\rm cm}$.
This value strongly depends on 
 $H_{\rm 0}$ ($r_{\rm E} \propto H_{\rm 0}^{-1/2}$) and 
 the lens mass ($r_{\rm E} \propto M_{\rm lens}^{1/2}$) 
 rather than the cosmological parameters in this case
 ($\sim 10\%$ uncertainty covers roughly all the reasonable range).
And finally, we get $10^{17} {\rm cm} \sim 7 \times 10^3 {\rm AU}$
 as the resultant upper limit to the source size.
This value is consistent with the result of the statistical research 
 performed by Wyithe et al. (2000c).
Moreover, for the $\sim 0.1M_{\odot}$ lens object case suggested 
 by Wyithe et al. (2000b) as a mean lens mass,  
 the size will be reduced by some factor 
 and become $\sim 2 \times 10^3 {\rm AU}$ !.
Alternatively, the lens object may be a stellar object, 
 and so, there is an upper mass limit to exist stably.
Even if we adopt this upper limit $\sim 100M_{\odot}$ for the lens mass, 
 the size will be $\sim 0.3 {\rm pc}$, at most.
Therefore, our result strongly supports the existence of an accretion disk 
 in a quasar and that the accretion disk smaller than this size 
 is a fairly dominant source of radiation from the quasar, 
 at least in the {\it V}-band at observer frame.
Additionally, resultant effective transverse velocity 
 on the source plane is $\lsim 10^5 {\rm (km~s^{-1})}$, 
 and is also consistent to the value presented by 
 Wyithe, Webster \& Turner, 1999. 

On the other hands, if we assume that the accretion disk is 
 a type described by the standard accretion disk model 
 (Shakura \& Sunyaev 1973), 
 we can also estimate the effective source size from its luminosity.
The magnitude of this quasar in the absence of a macrolens effect is 
 easily converted into flux, $f_{\nu}$ (Wozniak et al. 2000a).
If we denote the absorption as $A_{\rm V} {\rm ~mag}$
 and adopt the luminosity distance ($d_{\rm L}$) to the quasar, 
 luminosity of the quasar at this observed waveband ($L$)
 will estimated to be 
 $L \sim \nu (f_{\nu} \cdot 10^{0.4A_{\rm V}}) 4 \pi d_{\rm L}^2 
 \sim 3.7 \times 10^{42+0.4A_{\rm V}} {\rm ~erg~s^{-1}}$.
Furthermore, radiation process of the standard accretion disk is
 blackbody radiation, and the effective temperature ($T_{\rm eff}$) 
 of the accretion disk at this waveband corresponds to 
 $\sim 2.5 \times 10^4 ~{\rm K}$. 
On the other hands, we can relate the effective temperature, 
 luminosity and the radius 
 through the central black hole mass and accretion rate, 
\begin{equation}
T_{\rm eff} \sim \left( \frac{3}{4\pi \sigma} \right)^{1/4}
 L^{1/4} r^{-1/2},
\label{eq:disk}
\end{equation}
 where, $\sigma$ is the Thomson-scattering cross section.
Consequently, we are able to estimate the effective source size being  
 $ r \sim 2.0 \times 10^{14+0.2A_{\rm V}} {\rm ~cm}$. 
There is an uncertainty in $A_{\rm V}$, but 
 this is consistent with our results 
 and strongly indicates the existence of an accretion disk. 
We should pay attention that we do not insist the existing source 
 should be a standard-type accretion disk 
 (other accretion disk models may be also consistent with our results).
Rauch \& Blandford (1991) reported that the accretion disk in Q2237+0305
 is non-thermal or optically thin.
But, there are some ambiguities in this work (e.g., absorption, lens mass),
 and it is quite difficult to support or oppose the report from our results.
To specify disk models, we have to do more extensive monitorings 
 and/or analyze multi-band microlens lightcurves 
 to compare the resultant source sizes.

The performed fitting procedure do work, but strictly speaking, 
 all our best-fit reduced $\chi^2$ are somewhat larger than $1$. 
This means the goodness of fit of our results is not extremely good
 and probably, there may be some systematic errors 
 that we do not take into account.
In this work, we neglect the detailed feature about source, 
 magnification patterns and intrinsic variabilities of the quasar.
However, if such effects are really existing, 
 the shape of the lightcurve will be systematically altered 
 and the best-fit reduced $\chi^2$ may be increased by these effects.
There are difficulties to be take into account all above possibilities
 in our procedure, but it will be done in future.

In future work, we should further develop 
 quasar microlens technique in two statistical ways.
One is to study statistical properties of magnification near FC and CC.
Although the statistical features have already been studied by 
 many researchers (e.g., Wambsganss \& Kundi\'c 1995), 
 the effect, such as, lens object clustering 
 that affect properties of magnification in the vicinity of caustic
 is not well understood.
The other is to continue monitoring this kind of quasars, 
 sample similar microlensing events more and more. 
Such analysis may be able to reduce ambiguities arising due to 
 unknown lens mass, and/or different features of caustic networks and so on.

\acknowledgements

The author would like to express his thanks to
S. Mineshige for his extensive supports,  
E.L. Turner, J.S.B. Wyithe, K. Ioka, K. Mitsuda, K. Yoshikawa, T. Takeuchi, 
M. Umemura, A. Burkert and Y. Suto for their helpful comments, and 
anonymous referee for his/her valuable suggestions and comments.
The author also acknowledge the OGLE team for making their monitoring data
 publically available.
This work was supported in part by the Japan Society for
the Promotion of Science (9852).

\begin{table}
\begin{center}
\begin{tabular}{|r|c|c|c|c|c|c|c|}\hline
 ~ & fold-1 & fold-2 & fold-3 & fold-4 & cusp-1 & cusp-2 \\ \hline
 angle & $0.00$ & $0.50$ & $1.00$ & $1.50$ & $0.32$ & $1.57$ \\ \hline
 best fit & FC1 & FC1 & FC2 & FC1 & CC2 & CC2 \\ \hline
 $R (10^{-1})$ & $1.78^{-0.18}_{+0.19}$ & $2.42_{-0.22}^{+0.23} $  
  & $8.50_{-3.73}^{+1.32} $ & $7.64_{-0.72}^{+0.86} $  
  & $2.14_{-0.54}^{+1.36} $ & $1.00_{-0.24}^{+0.80} $ \\ \hline
 $v_{\rm x} (10^{-2})$ & $0.30_{-0.06}^{+0.06}$ 
  & $0.42_{-0.08}^{+0.08} $ & $ 0 $ & $1.31_{-0.25}^{+0.23} $ 
  & $-0.04_{-0.03}^{+1.79} $ & $-0.02_{-0.02}^{+0.79} $ \\ \hline 
 $v_{\rm y} (10^{-2})$ & $0.38_{-0.04}^{+0.04}$ & $0.49_{-0.14}^{+0.07} $
  & $2.74_{-0.72}^{+0.73} $ & $1.57_{-0.13}^{+0.17} $  
  & $0.63_{-0.59}^{+0.10} $ & $0.30_{-0.30}^{+0.04} $ \\ \hline
 $T_{\rm 0}$ & $1348_{-3}^{+3}$ & $1351_{-4}^{+2}$ 
  & $1366_{-2}^{+2}$ & $1350_{-4}^{+2}$ & $1366_{-8}^{+348}$ 
  & $1366_{-8}^{+308}$ \\ \hline
 $d (10^{-1})$ & $0$ & $0$ & $6.38_{-0.84}^{+0.47}$ & $0$ 
  & $3.24_{-6.58}^{+57.3} $ & $1.10_{-2.33}^{+20.8} $ \\ \hline 
 $M_{\rm 0}$ & $9.05_{-2.89}^{+2.37}$ & $9.23_{-2.27}^{+1.91}$ 
  & $18.55_{-1.58}^{+1.12}$ & $9.03_{-2.31}^{+2.27}$ & 
  $20.06_{-5.83}^{+1.05}$ & $20.02_{-6.46}^{+1.09}$ \\ \hline
 $\dot{M}_{\rm 0} (10^{-3})$ & $-3.05_{-1.59}^{+1.90}$ 
  & $-3.12_{-1.26}^{+1.49}$ & $-8.82_{-0.68}^{+0.96}$ 
  & $-3.02_{-1.52}^{+1.52}$ & $-9.88_{-0.18}^{+3.76}$
  & $-9.86_{-0.20}^{+4.18}$ \\ \hline
 $\bar{\chi}^2$ & $1.46$ & $1.49$ & $1.48$ & $1.47$ & 
  $1.33$ & $1.30$ \\ \hline 
\end{tabular}
\end{center}
\caption{Best fit parameters for several possible HME 
(see figure~\ref{fig:sche}) and their reduced $\chi^2$ ($\bar{\chi}^2$).
All the length scales and time scale is normalized by 
$r_{\rm E}$ and one day.
The unit of velocity is the Einstein-ring radius divided by a day. 
For $T_{\rm 0}$, $T_{\rm 0}=0$ correspond to $JD-2450000 = 0$.
Upper and lower value denoted beside the best fit parameters 
 show a $90\%$ confidence level which calculated from Monte-Carlo simulation
 (see the text).}
\label{tab:wp}
\end{table}

\end{document}